\documentclass[pra,showpacs,11pt,onecolumn]{revtex4}%
\usepackage{amsfonts}
\usepackage{amsmath}
\usepackage{amssymb}
\usepackage{graphicx}%
\begin{document}
\title{Efficient and High-Fidelity Generation of Atomic Cluster State with Cavity-QED
and Linear Optics}
\author{X.L. Zhang$^{1,2}$}
\email{xlzhang168@gmail.com}
\author{K.L. Gao$^{1}$}
\email{klgao@wipm.ac.cn}
\author{M. Feng$^{1}$}
\email{mangfeng1968@yahoo.com}
\affiliation{$^{1}$State Key Laboratory of Magnetic Resonance and Atomic and Molecular
Physics, Wuhan Institute of Physics and Mathematics, Chinese Academy of
Sciences, Wuhan 430071, China}
\affiliation{$^{2}$Graduate School of the Chinese Academy of Sciences, Bejing 100049, China}


\begin{abstract}
We propose a scheme to generate cluster states of atomic qubits by using
cavity quantum electrodynamics (QED) and linear optics, in which each atom is
confined in a resonant optical cavity with two orthogonally polarized modes.
Our scheme is robust to imperfect factors such as dissipation, photon loss,
and detector inefficiency. Discussions are made for experimental feasibility
of our scheme.

\end{abstract}
\maketitle

Recently, much attention has been paid on entangled states for testing quantum
nonlocality \cite{1,2,3} and for achieving quantum information processing
with, for example, cavity QED \cite{4,5,6}, trapped ions \cite{7} and free
photons \cite{8}.

The focus of this work is on generation of cluster state \cite{9}, a special
multipartite entangled state essential to one-way quantum computing \cite{10}.
It is considered that this quantum computing idea opens up a new paradigm for
constructing reliable quantum computers by measurement
\cite{11,12,13,14,15,16}. We have noticed a recent experiment realizing
cluster states by photonic qubits \cite{16}. While for a one-way quantum
computing, the flying qubits are not good candidates in view of accurate
manipulation on quantum states. The same problem also exists for using flying
atoms \cite{13}. Anyway, the technique to manipulate individual photonic
polarizations is much ahead of that for atoms. Due to this fact, we may
consider static atoms combined with flying photons to carry out quantum
information processing. Based on this idea, Cho and Lee \cite{14} have
proposed a scheme to generate atomic cluster states through the cavity
input-output process, in which the atoms as static qubits are trapped in
cavities and a single photon is flying as an medium. While considering the
practical aspect regarding current single-photon technique, the success
probability of the scheme would be quite small due to uncontrollable
imperfection and the cascade characteristic may also limit the length of the
cluster state. Particularly, additional single-photon resources are required
in that scheme, which is also an experimental challenge. In another paper
\cite{17}, an efficient strategy using entangling operation to build cluster
states was proposed. However, for each of the entangling operations, a
double-heralded single-photon detection and twice rotations of static qubits
are required, which complicate the implementation and prolong the operational
time. This is not good particularly for the case of low success rate of that
scheme. In addition, generation of four-atom cluster states proposed in
\cite{18} is strongly sensitive to quantum noise and thereby is difficult to
be extended to many-atom cases.

In this Brief Report, we propose an alternative scheme to generate cluster
states using cavity decay and considering symmetric use of polarizing beam
splitters (PBSs) and photon detectors. The photons, emitted by the atoms,
leaking out of the cavities and passing through PBSs become entangled, and
then the detection by detectors could map the entanglement from the photons to
the atoms. The favorable features of our scheme include: (1) The static qubits
are in fully controllable atoms, and post-selection by detectors makes the
scheme robust to photon loss and other sources of error such as spontaneous
emission, mismatch of cavity parameters and detector inefficiency. Although
the dissipative factors reduce the success rate of our scheme, the fidelity of
the generated cluster state is not affected; (2) The scheme is easily
restarted. So in the absence of dark counts of the detectors, we may achieve
many-atom cluster states with simpler and faster operations than in previous proposals,

For convenience of our description, we will focus on four-qubit cluster states
in most part of the paper. Consider an atom, with three degenerate excited
states and three degenerate ground states as shown in Fig. 1, where we
consider $(1,-1)=\left\vert g\right\rangle $, $(1,1)=\left\vert e\right\rangle
$ in the ground states to be qubit states and $(1,0)=\left\vert \alpha
\right\rangle $ in the excited states to be an ancillary state. The
transitions $\left\vert \alpha\right\rangle \rightarrow\left\vert
g\right\rangle $ and $\left\vert \alpha\right\rangle \rightarrow\left\vert
e\right\rangle $ could be coupled by left- and right-polarized radiation,
respectively. Suppose that we have four such atoms with each confined in an
optical cavity, as shown in Fig. 2. The cavities are identical and each of
them has two orthogonally polarized modes to resonantly couple $\left\vert
\alpha\right\rangle $ to $\left\vert g\right\rangle ,$ and to $\left\vert
e\right\rangle ,$ respectively. As each cavity is one-sided, the photons
leaking away will reach the PBSs as we design in Fig. 2. We assume the
cavities to be initially vacuum and each atom to be initially in the state
$\left\vert \alpha\right\rangle $, the latter of which could be achieved by a
pumping from the ground state $\left\vert \alpha^{\prime}\right\rangle $ by a
resonant $\pi$-polarized laser pulse before the scheme gets started. So the
initial state of the whole system could be written as $\left\vert
\psi(0)\right\rangle =%
{\textstyle\prod\limits_{k=1}^{4}}
\left\vert \alpha,0_{L},0_{R}\right\rangle _{k}$, where $\left\vert
\cdots\right\rangle _{k}$ denotes the atomic state, the left- and
right-polarized modes of the optical cavity k, respectively. In the
interaction picture and under the rotating-wave approximation, the Hamiltonian
in the kth (k=1, 2, 3, 4) cavity is (assuming $\hbar=1$) \cite{19},%
\begin{equation}
H_{k}=\frac{h_{k}}{2}\left[  a_{k,L}^{+}\left(  \left\vert g\right\rangle
_{kk}\left\langle \alpha\right\vert +\left\vert a^{\prime}\right\rangle
_{kk}\left\langle e^{\prime}\right\vert \right)  +a_{kR}^{+}\left(  \left\vert
e\right\rangle _{kk}\left\langle \alpha\right\vert +\left\vert a^{\prime
}\right\rangle _{kk}\left\langle g^{\prime}\right\vert \right)  +h.c.\right]
,
\end{equation}
where $a_{kL}^{+}$ and $a_{k,R}^{+}$ create left- and right-polarized photons,
respectively, in the kth cavity. For simplicity, we have assumed that the
coupling strengths between the cavity modes and their corresponding trapped
atoms (i.e., for transitions $\left\vert \alpha\right\rangle \rightarrow
\left\vert g\right\rangle $, $\left\vert \alpha\right\rangle \rightarrow
\left\vert e\right\rangle $ and $\left\vert g^{\prime}\right\rangle
\rightarrow\left\vert \alpha^{\prime}\right\rangle $, $\left\vert e^{\prime
}\right\rangle \rightarrow\left\vert \alpha^{\prime}\right\rangle )$ have the
constant value $h_{k}$. This can be reached by setting the relevant
Clebsch-Gordan coefficients to be $C_{\alpha,g}=C_{\alpha,e}=C_{g^{\prime
},\alpha^{\prime}}=C_{e^{\prime},\alpha^{\prime}}=\frac{1}{\sqrt{2}}$.
Considering weak cavity decay and weak spontaneous emission from the excited
states under the condition that no dissipation actually occurs during our
implementation of the scheme, we may describe the system governed by a non-
Hermitian Hamiltonian as follows (assuming $\hbar=1$),%
\begin{align}
H_{k}  &  =\frac{h_{k}}{2}\left[  a_{k,L}^{+}\left(  \left\vert g\right\rangle
_{kk}\left\langle \alpha\right\vert +\left\vert a^{\prime}\right\rangle
_{kk}\left\langle e^{\prime}\right\vert \right)  +a_{kR}^{+}\left(  \left\vert
e\right\rangle _{kk}\left\langle \alpha\right\vert +\left\vert a^{\prime
}\right\rangle _{kk}\left\langle g^{\prime}\right\vert \right)  +h.c.\right]
\nonumber\\
&  -i\frac{\gamma}{2}\left(  \left\vert a\right\rangle _{kk}\left\langle
a\right\vert +\left\vert g^{\prime}\right\rangle _{kk}\left\langle g^{\prime
}\right\vert +\left\vert e^{\prime}\right\rangle _{kk}\left\langle e^{\prime
}\right\vert \right)  -i\kappa(a_{kL}^{+}a_{kL}+a_{kR}^{+}a_{kR}),
\end{align}
where $\gamma$ is regarding the spontaneous emission from the excited states
and 2$\kappa$ accounts for one side decay rate of the kth cavity. For
simplicity, we have assumed the same rates regarding spontaneous emissions
from different excited levels and the same decay rate for each mode of the
cavities. This assumption is for reaching maximum implementation efficiency of
our scheme discussed below, because our scheme would be affected by
differently shaped wavepackets in the case of different $\gamma$ and $\kappa$
for different atoms and cavities. After an evolution time $t$ from the initial
state $\left\vert \psi(0)\right\rangle =%
{\textstyle\prod\limits_{k=1}^{4}}
\left\vert \alpha,0_{L},0_{R}\right\rangle _{k},$ the system evolves to an
entangled state,%
\begin{align}
\left\vert \psi(t)\right\rangle  &  =%
{\textstyle\prod\limits_{k=1}^{4}}
\exp(-\frac{\kappa+\frac{\gamma}{2}}{2}t)\left[  \left(  \frac{(\kappa
-\frac{\gamma}{2})}{\beta}\frac{e^{\beta t}-e^{-\beta t}}{2}+\frac{e^{\beta
t}+e^{-\beta t}}{2}\right)  \left\vert \alpha,0_{L},0_{R}\right\rangle
_{k}\right. \nonumber\\
&  -\left.  i\frac{h_{k}}{\beta}\frac{e^{\beta t}-e^{-\beta t}}{2}\left(
\left\vert g,0_{L},1_{R}\right\rangle _{k}+\left\vert e,1_{L},0_{R}%
\right\rangle _{k}\right)  \right]  ,
\end{align}
with
\[
\beta=\frac{1}{2}\sqrt{\left(  \kappa+\frac{\gamma}{2}\right)  ^{2}-2\left(
\gamma\kappa+h_{k}^{2}\right)  }.
\]
So due to dissipative factors, a left-polarized or right-polarized photon is
created with the success probability
\begin{equation}
P_{k}=\exp\left[  -\left(  \kappa+\frac{\gamma}{2}\right)  t\right]  \left(
h_{k}\frac{e^{\beta t}-e^{-\beta t}}{2\sqrt{2}\beta}\right)  ^{2}.
\end{equation}
Once the deexcitation actually happens, before reaching the PBSs 1 and 2 as
shown in Fig. 2, each of the photons has to pass through a quarter-wave plate
(QWP), which transforms left- and right-polarized photons to be horizontally
($H$) and vertically ($V$) polarized, respectively. Thus the whole system
reaches the state,%
\[
\left\vert \psi_{1}\right\rangle =\frac{1}{4}%
{\displaystyle\bigotimes\limits_{k=1^{\prime}}^{4^{\prime}}}
(\left\vert g\right\rangle _{k}\left\vert H\right\rangle _{k}+\left\vert
e\right\rangle _{k}\left\vert V\right\rangle _{k}),
\]
where the subscripts 1', 2', 3', and 4' label positions after the action of
QWP as shown in Fig. 2. We have dropped time in above equation because the
expression could be written formally irrelevant to time. As the PBS plays the
role of a parity check on the input photons, the detection of a photon at each
output port projects the state $\left\vert \psi_{1}\right\rangle $ into an
entangled state of the four atoms, which, including the actions\ by a
half-wave plate (HWP), is,%

\begin{equation}
\left\vert \psi_{2}\right\rangle =\frac{1}{2}\left(  \left\vert
gggg\right\rangle _{1234}+\left\vert eegg\right\rangle _{1234}+\left\vert
ggee\right\rangle _{1234}-\left\vert eeee\right\rangle _{1234}\right)  .
\end{equation}
This means that, once we have a click in each of the detectors D$_{\text{1}}$,
D$_{\text{2}}$, D$_{\text{3}}$, and D$_{\text{4}},$ we have generated the
cluster state of the four atoms. Please note that the state in Eq. (5) is
actually equivalent, under a Hadamard transformation $H=\frac{1}{\sqrt{2}%
}\left(  \sigma_{x}+\sigma_{z}\right)  $ on the first and the last atoms, to
the cluster state defined in \cite{9} with $N=4$, i.e., $\left\vert \phi
_{4}\right\rangle =$ $\frac{1}{4}\overset{4}{\underset{a=1}{\otimes}}\left(
\left\vert e\right\rangle _{a}\sigma_{z}^{\left(  a+1\right)  }+\left\vert
g\right\rangle _{a}\right)  $.

If we don't have the click in each of the four detectors during a waiting
time, e.g., 3/$\kappa$, which means failure of our implementation, we have to
restart the scheme. This could be done by steps as follows: (1) Excite
$\left\vert g\right\rangle $ to $\left\vert g^{\prime}\right\rangle $ or
$\left\vert e\right\rangle $ to $\left\vert e^{\prime}\right\rangle $ by a
$\pi-$polarized laser pulse; (2) The dissipation induced by the cavity modes
yields both $\left\vert g^{\prime}\right\rangle $ and $\left\vert e^{\prime
}\right\rangle $ to $\left\vert \alpha^{\prime}\right\rangle $; (3) Excite
$\left\vert \alpha^{\prime}\right\rangle $ to $\left\vert \alpha\right\rangle
$ by a $\pi-$polarized laser pulse. Then our scheme is ready to be done again.
Actually, the above steps are also useful for fusing two cluster states into a
bigger one. Consider two independent four-qubit cluster states described above
are located in dot-dashed boxes, respectively, in Fig. 3. We first perform the
transformation $\left\vert g\right\rangle \rightarrow\left\vert g^{\prime
}\right\rangle $ and $\left\vert e\right\rangle \rightarrow\left\vert
e^{\prime}\right\rangle $ on the last qubit of one of the cluster states
(e.g., in block I labelled in Fig. 3) and on the first qubit in another (e.g.,
in block II in Fig. 3). Before both the detectors D$_{I}$ and D$_{II}$ click,
the total state of the system is
\begin{align}
\left\vert \widetilde{\psi}_{1}\right\rangle _{I,II}  &  =\frac{1}{4}\left(
\left\vert ggg\alpha^{\prime}\right\rangle _{I}|V\rangle_{I}+\left\vert
eeg\alpha^{\prime}\right\rangle _{I}|V\rangle_{I}+\left\vert gge\alpha
^{\prime}\rangle_{I}|H\right\rangle _{I}-\left\vert eee\alpha^{\prime}%
\rangle_{I}|H\right\rangle _{I}\right) \nonumber\\
&  \otimes\left(  \left\vert \alpha^{\prime}ggg\right\rangle _{II}%
|V\rangle_{II}+\left\vert \alpha^{\prime}egg\right\rangle _{II}|H\rangle
_{II}+\left\vert \alpha^{\prime}gee\right\rangle _{II}|V\rangle_{II}%
-\left\vert \alpha^{\prime}eee\right\rangle _{II}|H\rangle_{II}\right)  ,
\end{align}
where we have considered the action of QWP. After both detectors are fired, we
reach a six-atom entangled state,%

\begin{align}
\left\vert \widetilde{\psi}_{2}\right\rangle _{I,II}  &  =\frac{1}{2\sqrt{2}%
}(\left\vert gggggg\right\rangle +|eegggg\rangle+|ggggee\rangle+|eeggee\rangle
\nonumber\\
&  +\left\vert ggeegg\right\rangle -\left\vert eeeegg\right\rangle -\left\vert
ggeeee\right\rangle +\left\vert eeeeee\right\rangle ),
\end{align}
where we have omitted the product states $\left\vert \alpha^{\prime
}\right\rangle $ regarding the last atom of the cluster state I and the first
atom of the cluster state II. So the entangled state only exits in six atoms.
By carrying out a Hadamard transformation on the first and last ones of the
six atoms, respectively, we get to a cluster state with six atoms. By this way
we can generate many-qubit cluster state, for example, of length N+M-2 from
two cluster states of respective lengths N and M.

We now give a brief discussion about the experimental matters of our scheme.
The level configuration under our consideration in Fig. 1 can be found in
$^{\text{87}}$Rb or $^{\text{171}}$Yb$^{\text{+}}$, for example, the level
with F=1 (e.g., 5$^{2}$S$_{1/2}$ of $^{\text{87}}$Rb or 6$^{2}$S$_{1/2}$ of
$^{\text{171}}$Yb$^{\text{+}}$) acts as the ground state and the excited state
could be 5$^{2}$P$_{3/2}$ of $^{\text{87}}$Rb or 6$^{2}$P$_{1/2}$ of
$^{\text{171}}$Yb$^{\text{+}}$. We consider $^{\text{87}}$Rb confined in an
optical cavity as Cs in \cite{20}. Although the atom is moving, as it moves
much slowly with respect to the photon, and it is well controllable, the atom
can be considered as a good carrier of static qubits. Alternatively, we may
suppose the atom $^{\text{87}}$Rb to be fixed by an optical lattice embedded
in a cavity, as done in \cite{21} for an ensemble of $^{\text{87}}$Rb. Using
the numbers in \cite{21} with the coupling strength $h_{k}=2\pi\times27$ MHz
and the cavity decay rate $\kappa=2\pi\times2.4$ MHz, and supposing the atomic
decay rate $\gamma=2\pi\times6$ MHz, we have the success probability 0.208 for
all the four cavities to emit photons simultaneously. Moreover, we may employ
$^{\text{171}}$Yb$^{\text{+}}$ in an ion trap which is embedded in an optical
cavity. Such a case has been demonstrated experimentally using Ca$^{+}$
\cite{22}$.$ If we assume the coupling strength to be $h_{k}\sim30$ MHz and
the cavity decay rate $\kappa\ $to be about 10 times smaller than the coupling
strength \cite{23}, we have the success probability of our scheme larger than
0.16 in the case of the atomic decay rate $\gamma<10$ MHz. As $^{\text{171}}%
$Yb$^{\text{+}}$ could be well localized in individual ion traps for a long
time, such a system is very suitable for our job. We expect in future
experiments higher Q cavities, lower spontaneous emission rate, and larger
cavity-atom coupling to increase the success rate of our scheme.

Since the quantum logical operation with photons is basically probabilistic,
when the photons go through each PBS, the success probability would decrease
by one half. As a result, a cluster state of four atoms is obtained in an
ideal implementation of our scheme only with the success rate 1/ 8, which is
the same as in \cite{17}. However,\ compared to \cite{17} with additional
operations necessary on the atoms, our scheme is much simpler in the many-atom
case. Moreover, we may compare with \cite{14}, a previous scheme to generate
cluster state with sequential single-photon interaction with different
cavities. The success probability of that scheme is proportional to
$(1-\eta)^{2n}$, with $\eta$ the loss rate of the single-photon, and $n$ the
number of components of the cluster state. In contrast, in the case of large
loss rate $\eta,$ e.g., the large inefficiency of the fiber, our scheme is
more efficient with the success probability proportional to $(1-\eta)^{n}$.
Moreover, the implementation of our scheme is faster due to photons output
from the cavities in parallel. In addition, better than \cite{18}, our scheme
can be directly extended to the preparation of cluster states of any size, or
two- or three-dimensional cluster states, which are prerequisites of a
meaningful quantum computation. However, to achieve a scalable scheme the
effect of dark counts of the detectors should be seriously considered. For a
normal dark count rate 100 Hz, the dark count probability for single photon in
our scheme is estimated to be $10^{-5},$ which would be significant in the
case of thousands detections. We hope the future technical advance could
overcome this difficulty.

In summary, we have presented a scheme to generate atomic cluster states by
using cavity QED, linear optical elements and photon detection. The distinct
advantages of our scheme are that the fidelity of the generated state is
insensitive to the quantum noise and the detection inefficiency, and the
scheme is easily restarted and in principle scalable. Therefore despite the
imperfect factors, the relaxation of experimental requirement makes our scheme
achievable with current techniques.

This work is supported in part by National Natural Science Foundation
of\ China under Grant Nos. 10474118 and 60490280, by Hubei Provincial Funding
for Distinguished Young Scholars, and by the National Fundamental Research
Program of China under Grants No. 2005CB724502 and No. 2006CB921200.


\begin{thebibliography}{99}                                                                                               %


\bibitem {1}A. Einstein, B. Podolsky, and N. Rosen, Phys. Rev. \textbf{47},
777 (1935).

\bibitem {2}J.S. Bell, Physics Long Island City, N.Y. \textbf{1}, 195 (1965).

\bibitem {3}D.M. Greenberger, M. Horne, A. Shimony, and A. Zeilinger, Am. J.
Phys. \textbf{58}, 1131 (1990).

\bibitem {4}E. Hagley, X. Ma\^{\i}tre, G. Nogues, C. Wunderlich, M. Brune,
J.M. Raimond, and S. Haroche, Phys. Rev. Lett. \textbf{79}, 1 (1997).

\bibitem {5}S. Bose, P.L. Knight, M.B. Plenio, and V. Vedral, Phys. Rev. Lett.
\textbf{83}, 5158 (1999).

\bibitem {6}S. Osnaghi, P. Bertet, A. Auffeves, P. Maioli, M. Brune, J.M.
Raimond, and S. Haroche, Phys. Rev. Lett. \textbf{87}, 037902 (2001).

\bibitem {7}Q.A. Turchette, C.S. Wood, B.E. King, C.J. Myatt, D. Leibfried,
W.M. Itano, C. Monroe, and D.J. Wineland, Phys. Rev. Lett. \textbf{81}, 3631 (1998).

\bibitem {8}P.G. Kwiat, K. Mattle, H. Weinfurter, A. Zeilinger, A.V.
Sergienko, and Y. Shih, Phys. Rev. Lett. \textbf{75}, 4337 (1995).

\bibitem {9}H.J. Briegel and R. Raussendorf, Phys. Rev. Lett. \textbf{86}, 910 (2001).

\bibitem {10}R. Raussendorf and H. J. Briegel, Phys. Rev. Lett. \textbf{86},
5188 (2001).

\bibitem {11}R. Raussendorf, D.E. Browne, and H.J. Briegel, Phys. Rev. A
\textbf{68}, 022312 (2003).

\bibitem {12}M.A. Nielsen, Phys. Rev. Lett. \textbf{93}, 040503 (2004).

\bibitem {13}X.B. Zou and W. Mathis, Phys. Rev. A \textbf{72}, 013809 (2005).

\bibitem {14}J. Cho and H.W. Lee, Phys. Rev. Lett. \textbf{95}, 160501 (2005).

\bibitem {15}D.E. Browne and T. Rudolph, Phys. Rev. Lett. \textbf{95}, 010501 (2005).

\bibitem {16}P. Walther, K.J. Resch, T. Rudolph, E. Schenck, H. Weinfurter, V.
Vedral, M. Aspelmeyer, A. Zellinger, Nature (London) \textbf{434}, 169 (2005).

\bibitem {17}S.D. Barrett and P. Kok, Phys. Rev. A \textbf{71}, 060310(R) (2005).

\bibitem {18}X.B. Zou, K. Pahlke, and W. Mathis, Phys. Rev. A \textbf{69},
052314 (2004).

\bibitem {19}M. Feng, Z.J. Deng, and K.L. Gao, Phys. Rev. A \textbf{72},
042333 (2005); D.L. Zhou, B. Sun, C.P. Sun, and L. You, Phys. Rev. A
\textbf{72}, 040302(R) (2005).

\bibitem {20}J. McKeever, J.R. Buck, A.D. Boozer, and H.J. Kimble, Phys. Rev.
Lett. \textbf{93}, 143601 (2004); A. Boca, R. Miller, K.M. Birnbaum, A.D.
Boozer, J. McKeever, and H.J. Kimble, Phys. Rev. Lett. \textbf{93}, 233603 (2004).

\bibitem {21}J.A. Sauer, K.M. Fortier, M.S. Chang, C.D. Hamley, and M.S.
Chapman, Phys. Rev. A \textbf{69}, 051804(R) (2004).

\bibitem {22}A.B. Mundt, A. Kreuter, C. Becher, D. Leibfried, J. Eschner, F.
Schmidt-Kaler, and R. Blatt, Phys. Rev. Lett \textbf{89}, 103001 (2002).

\bibitem {23}As the relevant data for $^{\text{171}}$Yb$^{\text{+}}$ in an
optical cavity is unavailable, we suppose these numbers for our calculation,
which do not change the physical essence of our discussion. Based on these
data, we want to show what prerequisites are necessary for achieving our scheme.

\textbf{Captions of the figures}

Fig. 1. The level configuration of the atoms. The dot-dashed and dashed lines
denote the coupling by left- and right- polarized cavity fields, respectively.

Fig. 2. The experimental setup for generation of a four-qubit cluster state.
The bold lines in the dashed box are four quarter-wave plates (QWP), which
transform left- and right-polarized photons to be horizontally and vertically
polarized, respectively. HWP means a half-wave plate working as a Hadamard
gate, i.e, $H\rightarrow\frac{1}{\sqrt{2}}\left(  H+V\right)  $ and
$V\rightarrow$ $\frac{1}{\sqrt{2}}\left(  H-V\right)  $. PBS is a polarized
beam splitter which transmits the state $\left\vert H\right\rangle $ and
reflects the state $\left\vert V\right\rangle $. $D_{i}$ (i=1, 2, 3, 4) are
single-photon detectors.

Fig. 3. The schematic for fusing two cluster states to be a larger one. In
each of the dot-dashed box there is a four-qubit cluster state. P represents a
unitary rotator for an operation $\sigma_{x}H\sigma_{x}=\frac{1}{\sqrt{2}%
}\left(  \sigma_{x}-\sigma_{z}\right)  $.
\end{thebibliography}
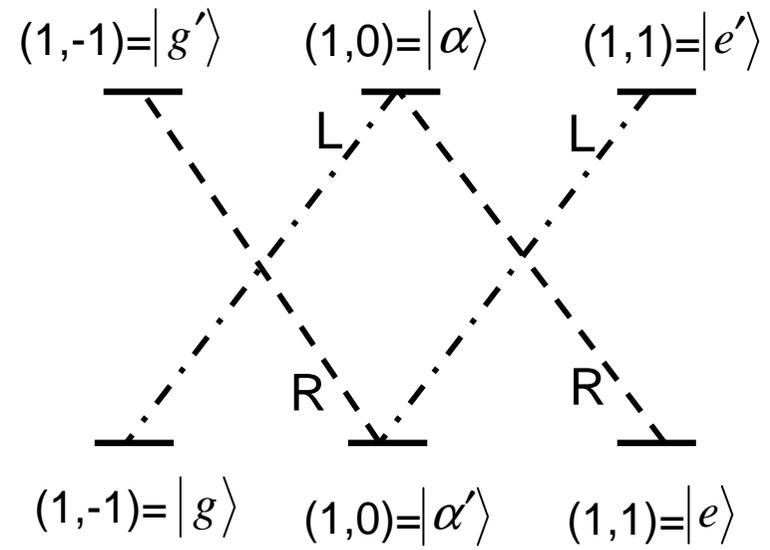
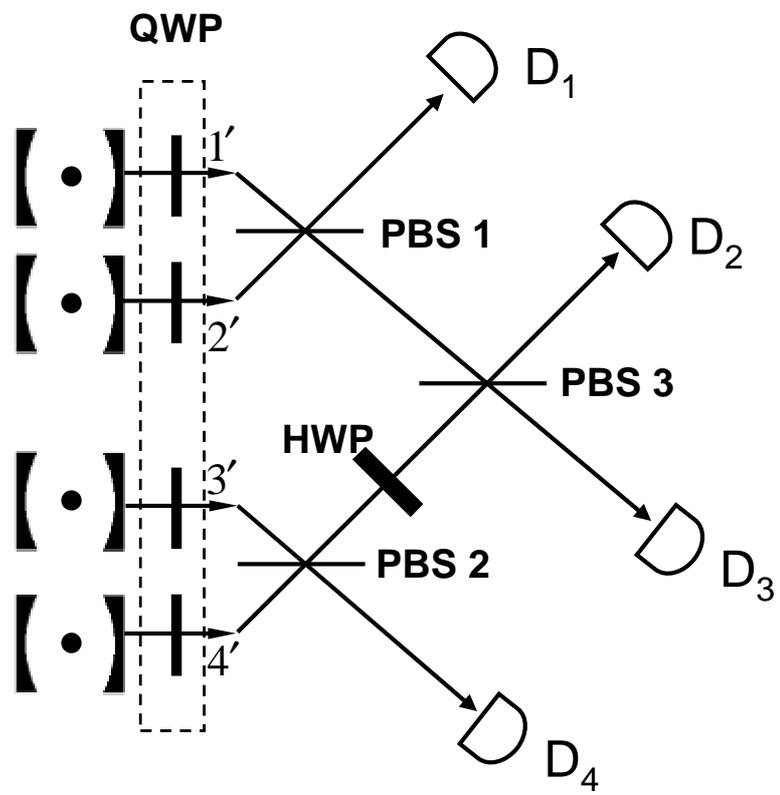
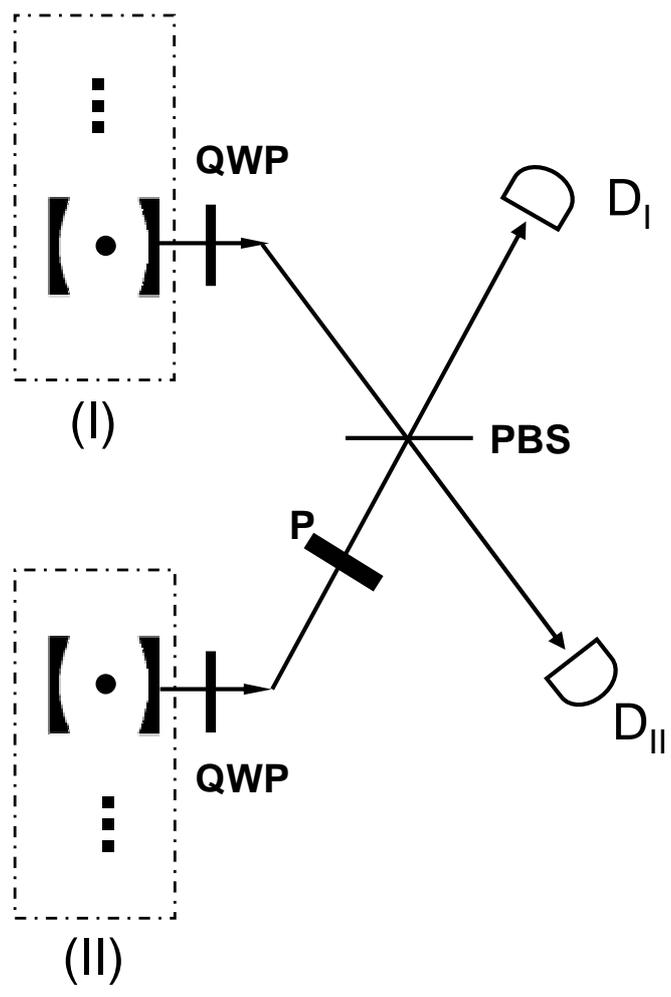
\end{document}